\documentclass{article}
\usepackage{ijcai26}

\usepackage{times}
\usepackage{soul}
\usepackage{url}
\usepackage[hidelinks]{hyperref}
\usepackage[utf8]{inputenc}
\usepackage[small]{caption}
\usepackage{graphicx}
\usepackage{amsmath}
\usepackage{amsthm}
\usepackage{booktabs}
\usepackage{algorithm}
\usepackage{algorithmic}
\usepackage[switch]{lineno}
\usepackage{amssymb}
\usepackage{booktabs}  % 三线表核心宏包
\usepackage{makecell}  % 处理单元格内多行内容
\usepackage{multirow}

\title{PhysFormer: A Physics-Embedded Generative Model for Physically Self-Consistent Spectral Synthesis}

\author{
	Siqi Wang$^1$, 
	Mengmeng Zhang$^1$, 
	Yude Bu$^{1,*}$,%\footnote{Corresponding author, E-mail: buyude@sdu.edu.cn},
	Chaozhou Mou$^2$
	\affiliations
	$^1$School of Mathematics and Statistics, Shandong University, Weihai, 264209, Shandong, China\\
	$^2$Weihai Institute for Interdisciplinary Research, Shandong University, Weihai 264209, Shandong, China
	\emails
	\{202417877, mmzhang\}@mail.sdu.edu.cn,\\
	\{buyude, mouchaozhou\}@sdu.edu.cn ($^*$Corresponding author: Yude Bu)
}

\begin{document}

\maketitle
%\footnotetext{Corresponding author: Yude Bu, E-mail: buyude@sdu.edu.cn.}
%\footnotetext[*]{Corresponding author: Yude Bu, E-mail: buyude@sdu.edu.cn.}

\begin{abstract}
    In scientific and engineering domains, modeling high-dimensional complex systems governed by partial differential equations (PDEs) remains challenging in terms of physical consistency and numerical stability. However, existing approaches, such as physics-informed neural networks (PINNs), typically rely on known physical fields or coefficients and enforce physical constraints via external loss functions, which can lead to training instability and make it difficult to handle high-dimensional or unobservable scenarios. To this end, we propose PhysFormer, a generative modeling framework that is self-consistent at both the data and physical levels. PhysFormer leverages a low-dimensional, physically interpretable latent space to learn key physical quantities directly from data without requiring known high-dimensional physical field parameters, and embeds the physical process of radiative flux generation within the network to ensure the physical consistency of the generated spectra. In high-dimensional, degenerate inversion tasks, PhysFormer constrains generation within physical limits and enhances spectral fidelity and inversion stability under varying signal-to-noise ratios (SNRs). More broadly, this approach shifts the physical processes from external loss functions into the generative mechanism itself, providing a physically consistent generative modeling paradigm for complex systems involving unknown or unobservable physical quantities.
    
\end{abstract}

\section{Introduction}

Accurately modeling and simulating complex systems is crucial in various scientific and engineering fields ~\cite{PeSANet[1]}. These systems are typically governed by partial differential equations (PDEs) and intricate physical processes, making it persistently challenging to achieve stable and scalable modeling while maintaining physical consistency. In recent years, physics-informed neural networks (PINNs) have sought to incorporate physical laws into deep learning models. Owing to their scalability ~\cite{meng2020ppinn[11]} and flexibility ~\cite{zhu2022neural[12]}, they have demonstrated promise in solving differential equations and inverse problems ~\cite{raissi2020hidden[13]}. However, most existing PINN-based methods mainly rely on external loss functions for physical constraints, making it difficult to automatically discover and embed physically consistent network structures ~\cite{nature}. Meanwhile, constructing physical loss functions typically requires known and explicitly parameterizable physical quantities, such as known equation coefficients, initial conditions or boundary conditions  ~\cite{raissi2019physics,sulishi}, which are not always available in real-world applications. As a result, the applicability of PINNs in more complex scenarios is limited.

Particularly in astrophysics, climate science, and plasma physics, many physical quantities required for solving governing equations—such as opacity, absorption coefficients—are often not directly observable. In such problems, conventional PINNs primarily act as equation solvers augmented with physical constraints. They struggle to function as data-driven models capable of discovering underlying physical structures. When such models are extended to ultra-high-dimensional output tasks, such as high-resolution spectral modeling, their limitations become increasingly pronounced. These issues manifest as unstable training and degraded generalization performance. In inverse problems, the models also tend to converge to local optima that satisfy PDE residuals but lack physical meaning ~\cite{India}.

Stellar spectral modeling provides a prototypical and highly challenging research scenario for studying such problems. The structure of stellar atmospheres is inherently constrained by established physical laws, such as the radiative transfer equation and the energy theorem. Mapping a small set of stellar parameters accurately to ultra-high-dimensional stellar spectra not only requires the model to have strong deep learning fitting capabilities but also to preserve the intrinsic physical consistency of the stellar atmosphere. Moreover, PINNs require the physical structure to be predefined. Therefore, they are not well suited for scenarios like stellar spectral modeling, where data are scarce and physical quantities must be inferred directly from the data.

To overcome these limitations, we propose a generative modeling framework that is self-consistent at both the data and physical levels. It extends the application of PINNs from equation solvers constrained by physics to physically consistent generative models. Rather than treating established physical laws as external constraints, physical processes are embedded directly into the network as part of its generative mechanism. This design does not rely on predefined PDE coefficients or scarce physical parameters. It can generate high-precision spectra that are consistent both numerically and with physical constraints.

We validate the effectiveness of the proposed framework on the PHOENIX theoretical spectral dataset spanning the ultraviolet to infrared bands, and the experimental results demonstrate that the method achieves superior stability, accuracy, and robustness in both high-dimensional spectral generation and parameter inversion tasks. In summary, we make the following contributions:
\begin{itemize}
	\item We propose a physically self-consistent framework at both data and physical levels, which is capable of learning representations of key physical quantities directly from data. It reduces the dependence of PINNs on fixed PDE coefficients and predefined physical parameters.
	
	\item PhysFormer embeds key physical processes, such as the radiative transfer equation and the energy theorem, directly into the network architecture. Instead of merely adding physical laws to the network loss function as external constraints, it enables physically consistent forward spectral generation.
	
	\item PhysFormer achieves high-precision, efficient modeling and parameter inference for ultra–high-dimensional stellar spectra. It provides a flexible spectral framework that is self-consistent at both the data and physical levels for stellar spectral modeling tasks.
\end{itemize}

\section{Related Work}
PINNs provide a general framework that integrates physical laws into the neural network training process as loss functions to solve PDE-driven forward and inverse problems ~\cite{2020Extended[15],2020Conservative[17],2020Physics[29]}. Existing methods can be broadly divided into two categories. The first category focuses on forward problems, aiming to approximate PDE solutions under fixed physical conditions with the associated physical fields and coefficients typically assumed to be known ~\cite{raissi2019physics}. The second category extends PINNs to inverse problems, in which some unknown parameters are treated as learnable variables and optimized jointly with the network weights to achieve parameter identification and system inversion ~\cite{yang2021b[7]}. Although PINNs have partially overcome the curse of dimensionality and are capable of approximating PDE solutions ~\cite{de2022error[31]}, most existing methods still rely on specifying the physical structure prior to training, which may limit their applicability in high-dimensional or partially unobservable physical fields.

To improve the data efficiency and physical consistency of PINNs, one class of methods attempts to automatically discover network structures through teacher-student network distillation ~\cite{nature}, and then extract physically meaningful structures through clustering and parameter reconstruction. However, such methods typically require substantial prior knowledge of governing equations, which limits their scalability in high-dimensional or very long sequence tasks. In addition, physics-encoded approaches directly embed physical constraints into the network architecture, such as PDE operators, conservation laws, and boundary conditions ~\cite{long2019pde,rao2023encoding,mi2024spatiotemporal,zeng2025phympgn}.

Stellar spectral modeling represents a highly challenging application scenario for physics-informed learning methods. Traditional template-matching methods are computationally expensive and difficult to scale. In recent years, deep learning approaches have achieved notable progress in this field. For example, ~\cite{[63]10.1093/mnras/staa3540} trained an encoder–decoder architecture on a large stellar spectral dataset for self-supervised spectral reconstruction. ~\cite{[65]li2022stellar} developed a deep convolutional neural network to predict key stellar atmospheric parameters. ~\cite{corral2023stellar[66]} trained deep learning models on a database of synthetic spectra. However, these approaches are primarily data-driven, and the learned features lack explicit physical interpretability. 

For forward generation and parameter inversion tasks, ~\cite{payne} proposed a framework that performs forward modeling from parameters to theoretical spectral fluxes and then conducts parameter inversion on observed spectra, but the model itself remains purely data-driven. In recent years, PINNs have also been partially applied to stellar spectral modeling. \cite{dahlbudding2024approximating} employed PINNs for radiative transfer modeling of exoplanetary atmospheres. ~\cite{kurucz} incorporated the hydrostatic equilibrium equation into the network architecture to generate stellar atmospheric structures. However, these methods rely on physical quantities provided by theoretical models and have not yet addressed the modeling challenges arising from unknown or unobservable physical quantities.

\section{Methodology}
\begin{figure*}[tb]  % [tb]：优先浮在页顶/页底（符合规范首选）
	\centering
	% 宽度设为\textwidth（跨双栏，符合“顶部/底部插图可跨双栏”的要求）
	\includegraphics[width=0.8\textwidth, keepaspectratio]{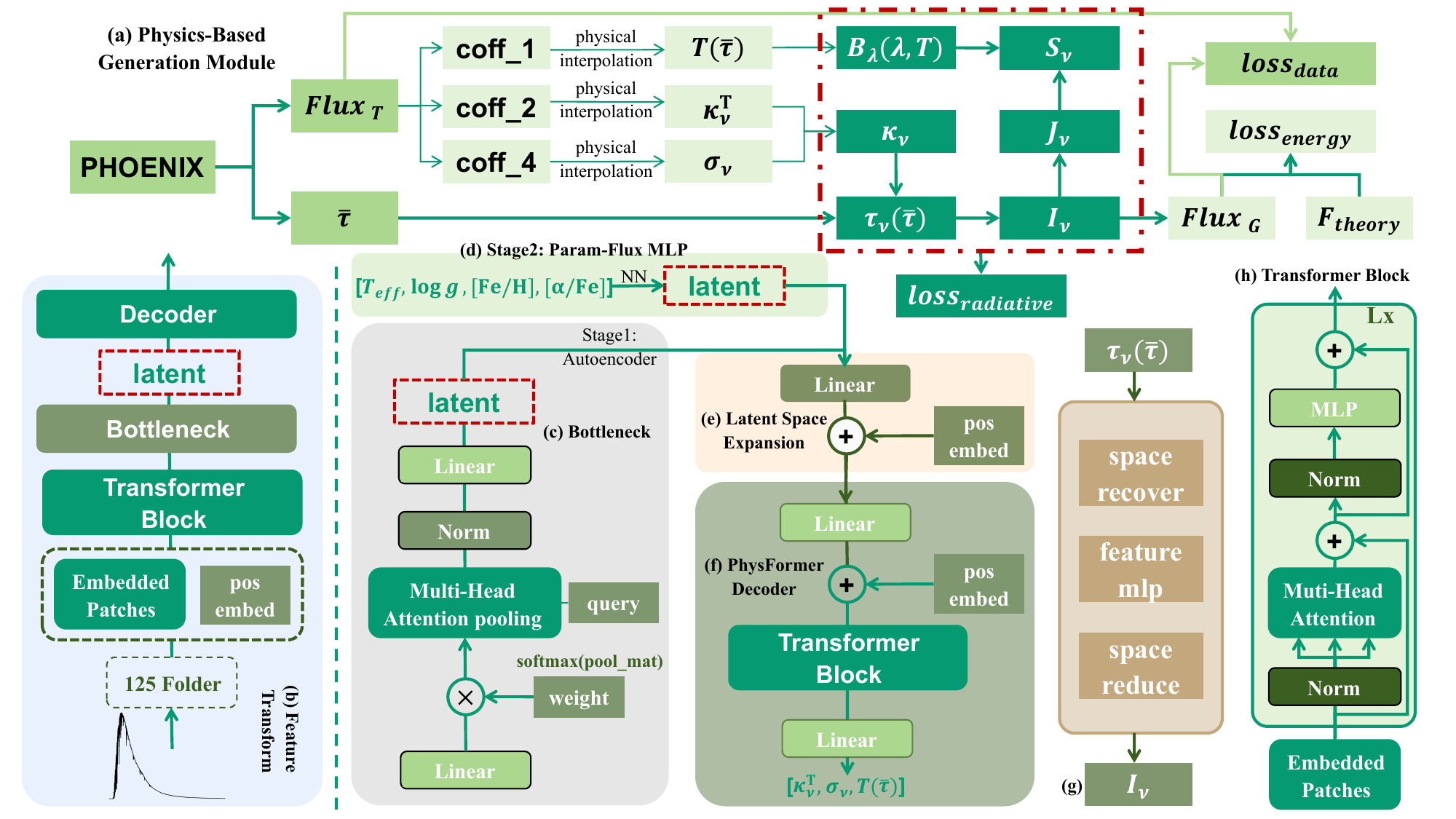}
	% 图例（caption）放在图下方，9号字体（模板自动适配）
	\caption{The architecture of PhysFormer. (a) Physical generation module that embeds physical processes directly into the network architecture. (b) Physically consistent spectral autoencoder. (c) Bottleneck architecture for generating a low-dimensional physical latent space. (d) Mapping network from stellar fundamental parameters to the physical latent space. (e) Latent Space Expansion Module. (f) PhysFormer decoder, used to generate key physical quantities required for subsequent physical computations. (g) Radiative Intensity Generation Module. (h) Transformer Block structure.}
	\label{fig:physformer_arch}  % 标签用于正文引用（如“See Figure \ref{fig:physformer_arch}”）
\end{figure*}

\subsection{Problem Formulation}
This paper focuses on a physics-constrained generative modeling and parameter inversion problem, with its application background rooted in stellar spectral modeling. The goal is, given a set of fundamental stellar atmospheric parameters
\begin{equation}
	\boldsymbol{\theta} = (T_{\mathrm{eff}}, \log g, [\mathrm{Fe/H}], [\alpha/\mathrm{Fe}]) \in \mathbb{R}^4,
\end{equation}
to generate the corresponding high-resolution stellar spectrum $\mathbf{F} \in \mathbb{R}^{W}$, covering an ultra-wide wavelength range across ultraviolet, optical, and infrared bands. The core challenge lies in the fact that stellar atmospheric structures are inherently constrained by the radiative transfer equation and the energy theorem, while key physical quantities—such as true absorption coefficients, scattering coefficients, and radiative intensity—are difficult to obtain directly. Consequently, the applicability of PINNs to high-dimensional spectral modeling and inversion tasks remains limited.

To this end, we propose a physically self-consistent generative modeling framework that learns physically meaningful latent representations directly from data, without assuming prior knowledge of physical field parameters, while explicitly embedding the physical generation process into the network architecture.
% 放在“首次讨论该图的段落之后”（不要放在文档末尾）

\subsection{PhysFormer Architecture}

PhysFormer (Figure \ref{fig:physformer_arch}) adopts a three-stage strategy: (1) physics-aware autoencoder pretraining, in which a bottleneck structure is introduced to learn a low-dimensional latent representation with physical interpretability; (2) parameter-to-latent mapping, where the spectral input is replaced by four-dimensional stellar parameters while the autoencoder is frozen; (3) parameter inversion, in which the trained PhysFormer is fixed and stellar parameters are optimized via $\chi^2$ minimization. Overall, the model can be regarded as a physics-embedded conditional generative model that learns latent physical fields. An embedded physical generator takes the latent representation as input and data-drivenly produces the intermediate physical quantities required by the radiative transfer equation and the energy theorem, followed by numerical integration to obtain the final spectrum. In this framework, physical laws are incorporated as integral components of the generative process rather than imposed as external loss terms, thereby promoting consistency between data fidelity and physical validity in the generated spectra.

\subsection{Physically consistent spectral autoencoder}
\subsubsection{Physics-Bottleneck Encoding Block}
In the encoder, this work adopts an architecture similar to Vision Transformer ~\cite{dosovitskiy2021an}, where the input spectrum is first partitioned into N non-overlapping patches along the spectral dimension. Multiple Transformer blocks are then applied for feature transformation to capture long-range dependencies across wavelengths. To mitigate the learning difficulties associated with directly mapping low-dimensional parameters to high-dimensional spectra, we introduce a token-level latent bottleneck to compress high-dimensional token sequences into a low-dimensional latent space with physically meaningful structure ~\cite{jaegle2021perceiver}:
\begin{equation}
	\mathbf{Z}_{tok}=\mathcal{B}(\mathbf{Z}_e),\quad \mathbf{Z}_{tok} \in \mathbb{R} ^{B\times K\times d_b},\quad \mathrm{K}\ll N.
\end{equation}
Here, $B(\cdot)$ denotes the bottleneck mapping operator, and $d_b$ represents the dimensionality of the latent-token mapping. Building on this, we draw on the core idea of $\beta-VAE$ ~\cite{betaVAE} to further compress the $K$ latent tokens into a one-dimensional global latent variable $Z_b \in \mathbb{R} ^{B\times l}$. This structure forces the model to encode full-spectrum information within a limited latent space, thereby facilitating the learning of stable and transferable physical representations that lay the foundation for subsequent parameter-driven generation.

\subsubsection{Physics-Decoded Block}
In the decoding stage, the low-dimensional latent representation is first dimensionally aligned with the number of patches in the original spectrum:
\begin{equation}
	\tilde{\mathbf{Z}}_{tok}=\mathcal{U} \left( \mathbf{Z}_b \right) \in \mathbb{R} ^{B\times N\times d_b},
\end{equation}
where $U(\cdot)$ denotes a token expansion operator implemented via a linear mapping. Subsequently, the expanded latent variables are projected into the decoder embedding space and augmented with positional encodings ($Z_d$). They are then processed by Transformer blocks and linearly mapped to generate high-resolution, wavelength-wise distributions of physical parameters required by the subsequent physics module, including true absorption coefficients, scattering coefficients, and temperature distributions:

\begin{equation}
	\{\boldsymbol{\kappa}, \boldsymbol{\sigma}, \boldsymbol{T}\} = \mathcal{H}(Z_d),
\end{equation}
where $\mathcal{H}(\cdot)$ denotes the physical decoding head. This process is consistent with the paradigm in neural operator methods, which generate high-dimensional physical fields from low-dimensional variables ~\cite{li2021fourier}. 

\subsubsection{Generative Physical Module}
\textbf{Physically Consistent Generative Spectral Modeling.} After obtaining data-driven intermediate physical quantities, PhysFormer generates spectra using differentiable physics-inspired operators. First, a physically motivated optical depth interpolation expands the low-dimensional parameters to the full optical depth, capturing their variation with depth. It then computes frequency-dependent optical depth, radiation intensity, and their angular averages, and finally generates the emergent spectral flux via numerical integration under the plane-parallel atmosphere assumption.

The propagation path of the optical depth is discretized into L optical layers $\{\boldsymbol{s}_i\}_{i=1}^{L}$, and the optical depth at a given frequency can be approximated in the form of a cumulative sum:
\begin{equation}
	\tau _{\nu ,i}=\sum_{k=1}^i{\chi _{\nu ,k}\Delta s_k,\qquad \Delta s_k=s_k-s_{k-1}.}
\end{equation}
Here, $\chi_\nu = \kappa_\nu + \sigma_\nu$, which is constrained to be non-negative.

The radiation intensity $I_\nu$ depends on both the optical depth $\tau$ and the directional cosine $\mu =\cos\theta$. We parameterize $I_\nu$ using a lightweight convolutional neural network, such that
\begin{equation}
	I_{\nu}(\tau _{\nu},\mu )\approx \mathcal{N} _I\left( \frac{\tau _{\nu}}{\mu} \right).
\end{equation}
Here, $\frac{\tau _{\nu}}{\mu}$ represents the effective optical depth when radiation propagates along different directions. This modeling approach preserves the physical dependency of radiation intensity on both optical depth and direction.

We discretize the angular integration using Gauss–Legendre quadrature ~\cite{press2007numerical}. $\left\{ \mu _i,w_i \right\} _{i=1}^{N_{\mu}}$ denote the discrete angular points and their corresponding quadrature weights, respectively. The generated flux $F_\nu$ can be approximated as:
\begin{equation}
	F_{\nu}=2\pi \int_{-1}^1{I_{\nu}\mu d\mu}.
\end{equation}

\textbf{Radiative Transfer Equation.} The study of radiative transfer is crucial in many scientific fields, including astrophysics ~\cite{mishra2021physics}. In this work, we introduce a differentiable RTE residual as a physics-based regularization term, encouraging the evolution of radiation intensity with optical depth to remain consistent with physical laws:
\begin{equation}
	\mathcal{R}_{\mathrm{RTE}} =\frac{d I_\nu}{d(\tau_\nu/\mu)}e^{-\frac{\tau _{\nu}}{\mu}}-I_\nu e^{-\frac{\tau _{\nu}}{\mu}}+S_\nu e^{-\frac{\tau _{\nu}}{\mu}}.
\end{equation}

Here, $S_\nu$ denotes the source function. The formulation penalizes physically inconsistent radiation transport by minimizing the RTE residual during training.

\textbf{Energy Theorem.} In stellar atmosphere modeling, it is necessary that the generated spectrum remains compatible with the total radiative energy within a given wavelength range based on the energy balance relation. We introduce an energy theorem residual based on the Stefan--Boltzmann law and Planck’s formula ~\cite{hubeny2014theory}:

\begin{equation}
	\mathcal{R}_{\mathrm{ET}} =\int_{\nu _1}^{\nu _2}{F_{\nu}\left( \nu ,T \right) d\nu -\sigma T^4}.
\end{equation}
Here, $\nu_1$ and $\nu_2$ represent the specific wavelength range.

These physics-based losses serve only as auxiliary regularization during training. The physical processes themselves are explicitly embedded in the forward generative pipeline rather than being imposed solely through loss functions.

\begin{figure*}[tb]  % [tb]：优先浮在页顶/页底（符合规范首选）
	\centering
	% 宽度设为\textwidth（跨双栏，符合“顶部/底部插图可跨双栏”的要求）
	\includegraphics[width=0.9\textwidth]{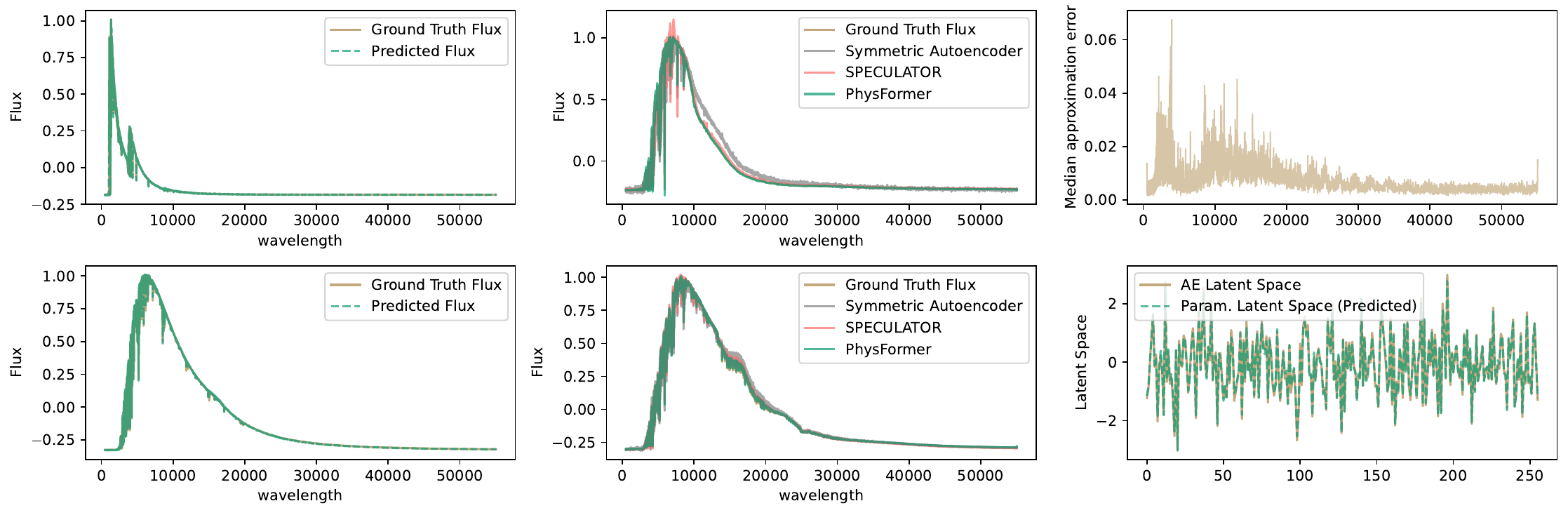}
	% 图例（caption）放在图下方，9号字体（模板自动适配）
	\caption{Visualization of Spectra Generated by PhysFormer. Left: comparison between PhysFormer-generated spectra and ground truth. Middle: comparison among PhysFormer, Symmetric Autoencoder, and SPECULATOR. Top right: median spectral error across wavelengths. Bottom right: visualization of the learned low-dimensional physical latent space.}
	\label{fig:SMP}  % 标签用于正文引用（如“See Figure \ref{fig:physformer_arch}”）
\end{figure*}

\subsubsection{Mapping from Parameters to the Physical Latent Space}
In the forward modeling stage, PhysFormer generates high-dimensional stellar spectra from low-dimensional stellar parameters. The stellar parameter vector is defined as
\begin{equation}
	\boldsymbol{p} = (T_{\mathrm{eff}}, \log g, [\mathrm{Fe}/\mathrm{H}], [\alpha/\mathrm{Fe}]).
\end{equation}
Given that the physics-aware autoencoder has learned a low-dimensional latent space encoding physical information, its parameters are frozen during this stage to maintain a consistent latent structure. A lightweight MLP is used to map parameters into this latent space, enabling parameter-conditioned spectral generation. This design alleviates the complexity of directly mapping low-dimensional parameters to high-dimensional spectra while preserving the physical structure and stability of the latent space.

\subsubsection{Physics-consistent stellar parameter inversion}
In the parameter inversion task, PhysFormer is regarded as an end-to-end differentiable physical forward model. To simulate observational conditions, Gaussian noise is added to the theoretical spectra, and stellar parameters $p$ are estimated by minimizing the $\chi^2$ objective:
\begin{equation}
	\mathcal{L} _{\chi ^2}(p)=\sum_{\nu =1}^W{\frac{\left( F_{\nu}^{obs}-\hat{F}_{\nu}(p) \right) ^2}{\sigma _{\nu}^{2}}.}
\end{equation}
Here, $ F_{\nu}^{obs}$ represents the simulated observed spectrum, $\hat{F}_{\nu}(p)$ denotes the physically consistent generated spectrum, and $\sigma _{\nu}^{2}$ is the variance of the added noise, determined by the specified signal-to-noise ratio (SNR).

During the inversion, the forward model parameters are fixed, and only the stellar parameters are optimized:
\begin{equation}
	p^*=\arg\min \limits_{p}\mathcal{L} _{\chi ^2}\bigl( \hat{F}(p),F^{obs} \bigr) +\mathcal{L} _{phys}.
\end{equation}
This formulation empirically yields stable inversion behavior under varying SNR conditions.

\section{Experiments}

\subsection{Experimental Setup}
This section details the experimental setup, including the dataset, model configuration, and physical module settings. All experiments are designed to evaluate the effect of structured physics embedding on generative performance and inverse stability.

\textbf{Datasets.} We use the PHOENIX stellar atmosphere model spectra, consisting of 26{,}747 spectra, each with length $9875$, covering the ultraviolet, optical, and infrared bands. All spectra are normalized and partitioned into $N=125$ non-overlapping wavelength patches, with each patch containing $79$ consecutive wavelength points.

\textbf{Model Configuration.} In the physics-aware autoencoder, the encoder and decoder depths are set to $12$ and $3$, respectively, and the dimension of the physical latent space is $256$.

\textbf{Physical module.} The stellar atmosphere is approximated as a semi-infinite plane-parallel layer. The optical depth is discretized into $64$ layers, and angular integration is performed using Gauss--Legendre quadrature with $20$ directions over the interval $[-1,1]$.

\begin{table}[htbp]
	\centering  % 表格整体在页面居中
	
	\begin{tabular}{cccc}  % 所有列设为居中对齐（原lccc→cccc）
		\toprule
		Model & RMSE & MAE & $R^2$ \\
		\midrule
		\textbf{PhysFormer (Ours)} & \textbf{0.0054} & \textbf{0.0033} & \textbf{0.9997} \\
		PhysGNN & 0.1959 & 0.1622 & 0.7216 \\
		Symmetric Autoencoder & 0.0200 & 0.0123 & 0.9956 \\  % Model列内容也居中换行
		SPECLATOR & 0.0106 & 0.0064 & 0.9985 \\
		The Payne & 0.0618 & 0.0357 & 0.9372 \\
		Kurucz-a1 & 0.0502 & 0.0204 & 0.9760 \\
		\bottomrule
	\end{tabular}
	\caption{Comparison of PhysFormer with Other Baselines}  % IJCAI风格的英文标题
	\label{tab:physformer_baseline}  % 符合顶会命名规范的标签
\end{table}

\subsection{Spectral Modeling Performance of PhysFormer}

As shown in Figure \ref{fig:SMP}, the mean squared error between PhysFormer and the theoretical spectra is $4.40 \times 10^{-5}$. The error distributions in the ultraviolet ($500$--$4000\,\AA$) and infrared ($7500$--$55{,}000\,\AA$) bands are relatively stable and remain overall low. In contrast, the errors in the optical band ($4000$--$7500\,\AA$) are slightly higher, with the median error peaking within this interval. Overall, PhysFormer demonstrates accurate and stable spectral reconstruction for the majority of characteristic spectral lines.

\begin{table*}[tb]
	\centering
	\small  % 表格内容9pt，适配宽度
	\begin{tabular*}{\textwidth}{@{\extracolsep{\fill}} ccccc @{}}  % 满宽+数值右对齐
		\toprule
		Model Configuration & \makecell{MSE(AE \\ 200epoch)} & MSE(P2F) & RTE Loss & ET Loss \\
		\midrule
		\textbf{PhysFormer (Ours)} & $\mathbf{6.24 \times 10^{-5}}$ & $\mathbf{4.41 \times 10^{-5}}$ & $\mathbf{5.16 \times 10^{-5}}$ & \textbf{0.0195} \\
		w/o Physics Process & $8.41 \times 10^{-5}$ & $4.28 \times 10^{-5}$ & - & - \\
		w/o Bottleneck & $3.75 \times 10^{-5}$ & 0.0489 & \makecell{$5.58 \times 10^{-5}$(AE)\\$3.09 \times 10^{-5}$(P2F)} & \makecell{0.0236(AE)\\0.1210(P2F)} \\
		Smaller Bottleneck & $8.60 \times 10^{-4}$ & - & $7.67 \times 10^{-5}$ & 0.0164 \\
		MLP Decoder & $2.83 \times 10^{-4}$ & - & $1.17 \times 10^{-4}$ & 0.0189 \\
		Reduced Angular Integration & $8.87 \times 10^{-4}$ & - & $2.66 \times 10^{-5}$ & 0.0160 \\
		\bottomrule
	\end{tabular*}
	\caption{\small Ablation Experiment Results. AE refers to the autoencoder stage results (pretrained for 200 epochs), and P2F refers to the results from the parameter-to-flux prediction stage. RTE Loss represents the mean squared error of the radiative transfer equation residuals, while ET Loss denotes the mean squared error of the energy theorem residuals. Due to the limited training epochs in the AE stage, its errors are generally higher than those in the P2F stage.}
	\label{tab:ablation}  % 启用Label，确保可引用
\end{table*}

\begin{figure*}[tb]  % [tb]：优先浮在页顶/页底（符合规范首选）
	\centering
	% 宽度设为\textwidth（跨双栏，符合“顶部/底部插图可跨双栏”的要求）
	\includegraphics[width=0.9\textwidth]{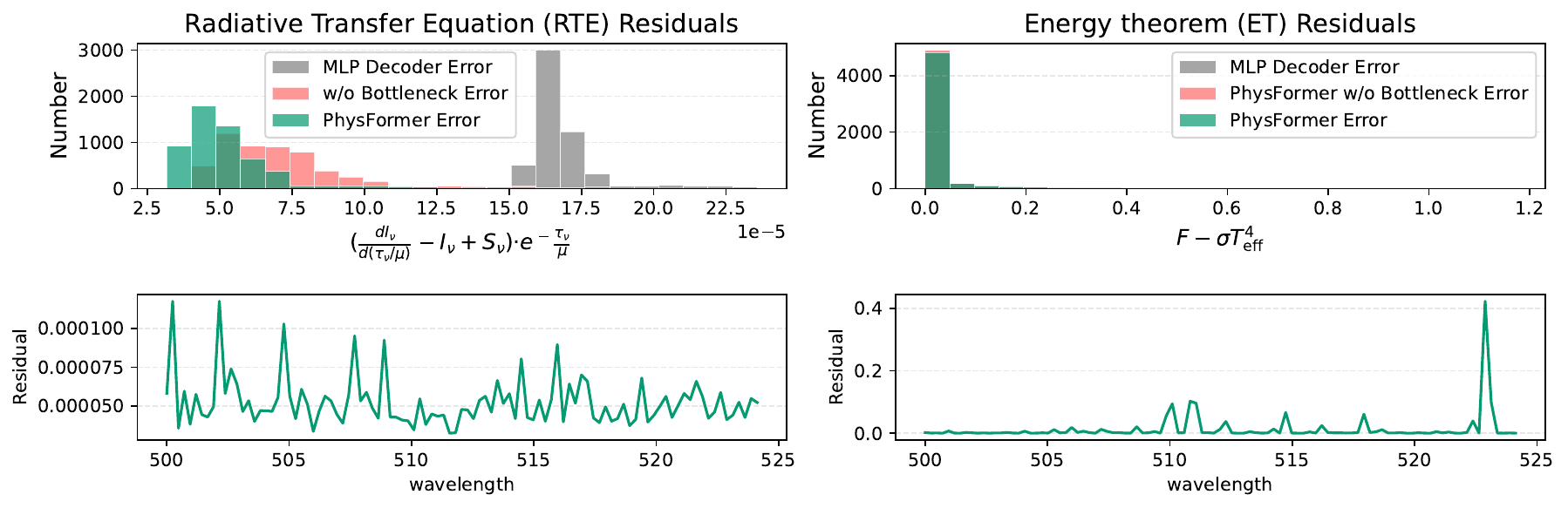}
	% 图例（caption）放在图下方，9号字体（模板自动适配）
	\caption{Residual Distributions of RTE and ET on the Test Set. Top row: Residual distributions of the Radiative Transfer Equation (RTE) and Energy Theorem (ET) for the three models: PhysFormer, PhysFormer without Bottleneck, and MLP Decoder. Bottom row: Absolute residual distributions of RTE and ET for the first 100 wavelength points for PhysFormer.}
	\label{fig:loss_distri}  % 标签用于正文引用（如“See Figure \ref{fig:physformer_arch}”）
\end{figure*}

\subsection{Comparison with Existing Models}
To comprehensively evaluate the effectiveness of PhysFormer, we selected several representative baseline methods that encompass different modeling approaches in high-dimensional scenarios.

\textbf{The Payne} ~\cite{payne}. A representative data-driven framework for stellar spectrum generation and parameter inversion, which employs a MLP to directly learn the mapping from stellar parameters to spectral flux, without explicitly incorporating physical structures or intermediate physical quantities.

\textbf{Kurucz-a1} ~\cite{kurucz}. We leverage its dual-encoder architecture to perform block-wise modeling of parameters and wavelength, and then concatenate the features to predict the spectral flux.

\textbf{PhysGNN} ~\cite{physgnn}. A physics-inspired graph neural network approach that represents physical variables as graph nodes and models their interactions via learned relational structures.

\textbf{Symmetric Autoencoder} ~\cite{gebran2024generating}. A symmetric autoencoder-based model that constructs a latent representation of stellar spectra and reconstructs the spectra through a mirrored decoder.

\textbf{SPECULATOR} ~\cite{alsing2020speculator}. A spectral emulation method that enables fast and accurate generation of galaxy spectra via PCA dimensionality reduction and neural network fitting.

We use three types of metrics to quantitatively assess the model performance: Root Mean Square Error (RMSE), Mean Absolute Error (MAE), and coefficient of determination ($R^2$). Table \ref{tab:physformer_baseline} and the middle panel of Figure \ref{fig:SMP} show that purely data-driven methods such as The Payne exhibit limited performance when generating ultra-high-dimensional, cross-band, high-resolution spectral fluxes from only four-dimensional stellar parameters. While SPECULATOR and symmetric autoencoders can capture the overall spectral shape, they exhibit poor stability and consistency in line-sensitive regions. In contrast, PhysFormer achieves lower prediction errors and more stable spectral responses under complex parameter variations.

\subsection{Ablation Study}
To assess the contribution of each component in PhysFormer, we perform ablation studies on the physical modules, latent structure, and numerical integration (Table ~\ref{tab:ablation}).

\textbf{PhysFormer w/o Physics Process.} In this variant, the physical generation module and its associated physical constraints are removed, and latent variables are directly decoded into spectra. Although this model still attains a mean squared error on the order of $1 \times 10^{-5}$ during both the autoencoder training stage and the parameter-to-spectrum prediction stage, it essentially degenerates into a purely data-driven model, with the learned representations primarily capturing empirical correlations rather than possessing explicit physical interpretability. This variant exhibits substantially larger parameter estimation errors in the inversion task compared to the full PhysFormer, as discussed in Section 4.5.

\textbf{PhysFormer w/o Bottleneck.} To 	verify the necessity of the low-dimensional latent space, we removed the Bottleneck module, directly connecting the encoder output to the decoder. Although the autoencoder reconstruction error is very low ($3.75 \times 10^{-5}$), the mean squared error increases sharply to 0.0489 during the parameter-to-spectrum prediction stage. This indicates that the four-dimensional stellar fundamental parameters are difficult to directly map to the ultra-high-dimensional stellar spectra, and the Bottleneck design plays a crucial role in balancing reconstruction capability with the physical interpretability of the latent space.

\textbf{Smaller Bottleneck.} We reduced the dimensionality of the low-dimensional latent space from 256 to 128. Experimental results indicate that excessive compression of the latent space constrains the model’s expressive capacity, causing the spectral reconstruction error of the autoencoder to increase by more than an order of magnitude. This indicates that selecting an appropriate latent space dimensionality is crucial for preserving physically meaningful variations and alleviating the difficulty of parameter-to-spectrum mapping.

\textbf{MLP Decoder.} The Transformer-style decoder was replaced with an MLP decoder. Although the model can capture the overall spectral structure, both prediction accuracy and generalization were inferior to the Transformer decoder, indicating the advantage of Transformer architectures in modeling long-range dependencies in high-dimensional spectral sequences.

\textbf{Reduced Angular Integration.} The number of Gaussian directions considered in the semi-infinite plane-parallel atmosphere was reduced from 20 to 4. The results show a significant decrease in spectral accuracy, highlighting the importance of sufficiently accurate numerical integration for maintaining spectral generation performance.

In addition, the RTE Loss and ET Loss not explicitly labeled in Table~\ref{tab:ablation} correspond to the autoencoder phase. All models containing physical modules could compute the physical equation losses and maintain them at low levels, indicating that the model design remains numerically consistent with the embedded physical constraints. Additional checks confirm that the true absorption and scattering coefficients remain non-negative, and the generated temperature distributions remain within the bounds of the PHOENIX theoretical spectra, suggesting that PhysFormer learns physically consistent representations of intermediate quantities. It should be noted that flux prediction accuracy remains the primary evaluation criterion. Models with a lower RTE Loss may still produce inaccurate emergent fluxes, which highlights the necessity of incorporating global physical constraints and physically structured latent representations.

Figure \ref{fig:loss_distri} illustrates that PhysFormer (Ours) achieves the lowest and most tightly distributed RTE residuals, whereas the MLP Decoder exhibits substantially larger residuals. Regarding the energy theorem, all three models show comparable performance, indicating that the ET loss is primarily governed by the physical module.

\subsection{Parameter Inversion Experiments}

\begin{figure}[tb]  % [tb]：优先浮在页顶/页底（符合规范首选）
	\centering
	% 宽度设为\textwidth（跨双栏，符合“顶部/底部插图可跨双栏”的要求）
	% 宽度改为单栏宽度（\columnwidth），0.8倍避免图片太满
	\includegraphics[width=0.7\columnwidth, keepaspectratio]{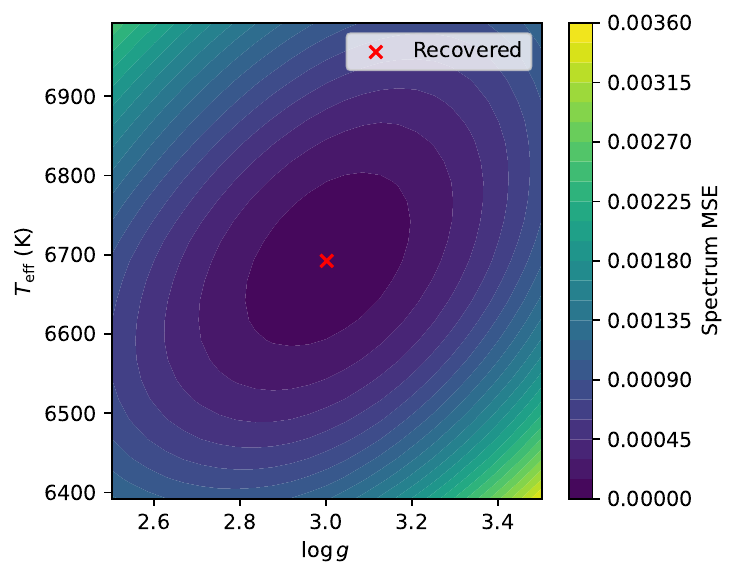}
	%\includegraphics[width=\textwidth]{fig5_degeneracy.pdf}
	% 图例（caption）放在图下方，9号字体（模板自动适配）
	\caption{Loss landscape in $T_{\mathrm{eff}}$--$\log g$ space for a single observed spectrum. Color indicates the spectrum reconstruction MSE. The red cross marks the recovered parameter. The elongated and tilted contours indicate a strong degeneracy between the two parameters.}
	\label{fig:degeneracy}  % 标签用于正文引用（如“See Figure \ref{fig:physformer_arch}”）
\end{figure}

In the parameter inversion experiments, we compared the performance of PhysFormer with PhysFormer w/o Physics Process under different SNR conditions. Unlike inversion studies that rely on strong priors or empirical templates, we consider a particularly challenging inversion setting characterized by: (1) low-SNR observations, where the inversion process must recover stellar parameters under significant noise interference; (2) The model relies solely on the spectra for parameter estimation, which significantly increases the risk of parameter degeneracy, i.e., the existence of distinct parameter combinations that produce nearly indistinguishable spectra (Figure \ref{fig:degeneracy}). This problem constitutes a highly nonlinear, strongly degenerate, and noise-sensitive inverse problem.

\begin{figure}[tbh!]  % 关键：去掉*，用[tbh!]提升当前位置优先级
	\centering
	% 宽度改为单栏宽度（\columnwidth），0.8倍避免图片太满
	\includegraphics[width=0.8\columnwidth, keepaspectratio]{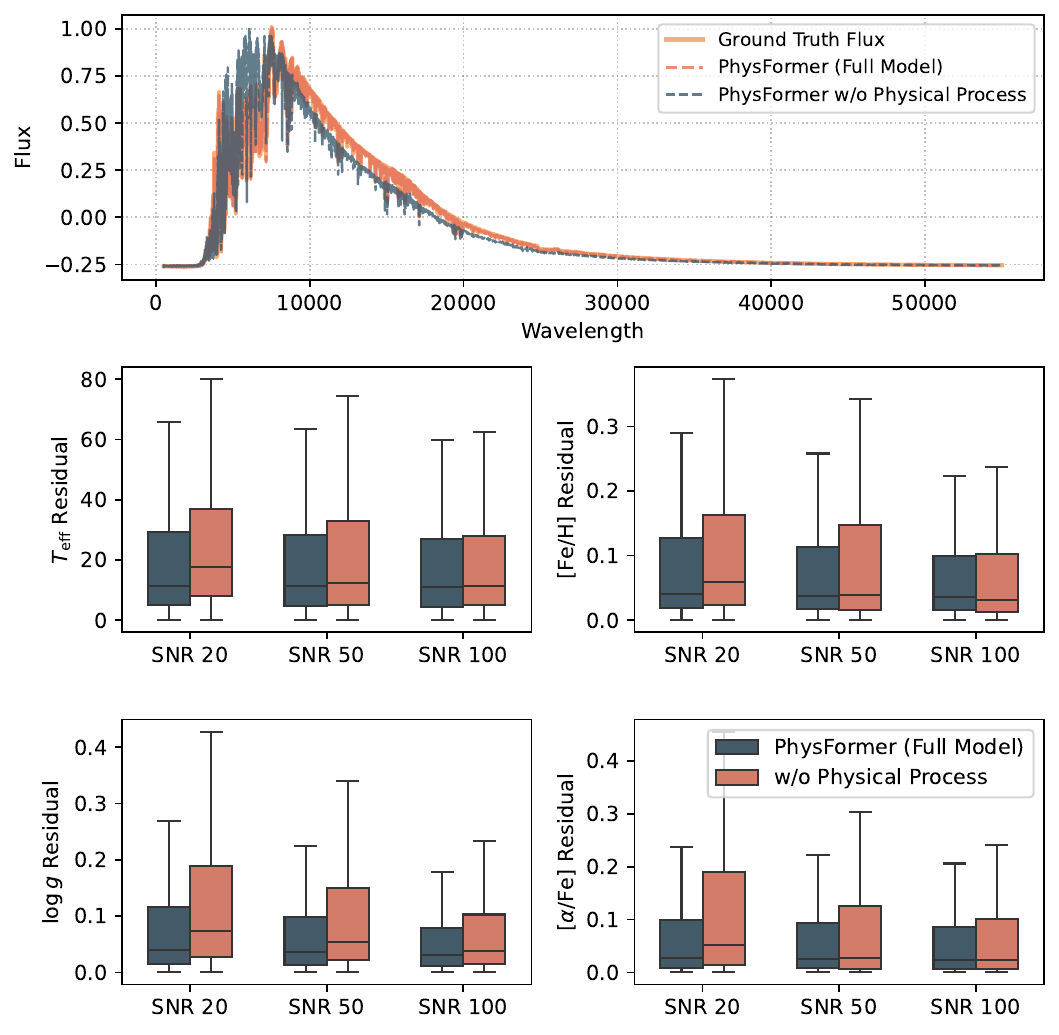}
	% 修正caption格式（完整句式，符合IJCAI学术规范）
	\caption{Example of Parameter Inversion Results and Error Distributions. Top: Example spectra generated from the inverted parameters, comparing the output of PhysFormer with PhysFormer w/o Physics Process. Bottom: Distributions of absolute errors in the inversion of four-dimensional stellar parameters under different signal-to-noise ratio (SNR) conditions for the two models.}
	\label{fig:box}  % label位置不变（仍在caption后）
\end{figure}

Figure \ref{fig:box} illustrates stellar parameter recovery by the two models. The top panel presents a rare star with solar metallicity and enhanced $\alpha$-elements $(T_{\mathrm{eff}} = 4300\,\mathrm{K},\ \log g = 5.5,\ [\mathrm{Fe}/\mathrm{H}] = 0.0,\ [\alpha/\mathrm{Fe}] = 1.2)$. For certain rare stars, PhysFormer maintains high inversion accuracy, producing spectra from the inferred parameters that more closely match the theoretical spectra. As shown in the bottom panel, the overall residual distribution of parameters inferred by PhysFormer is narrower and lower, and remains stable across different SNR conditions, indicating that the physically structured constraints contribute to improved robustness of the inversion process. Although all parameters remain challenging due to unavoidable degeneracies, PhysFormer significantly mitigates unstable inversion compared to baselines.
﻿

\section{Conclusion}
This work proposes PhysFormer, a physics-consistent, generative-modeling framework for stable spectral modeling and accurate parameter inversion, even when the underlying high-dimensional physical fields are unknown. Unlike conventional PINNs that primarily rely on external physics-based losses, PhysFormer directly embeds the physical generation process into the network and uses a low-dimensional, physically interpretable latent space to learn key radiative transfer quantities, producing predictions that are consistent at both data and physics levels. We validate its performance on an ultra-long stellar spectral dataset spanning ultraviolet, visible, and infrared bands with 9,875 wavelength points. Experiments show that PhysFormer achieves superior stability and consistency, even in spectrally sensitive regions. More broadly, by replacing the equations or constraints in the physics module while keeping the overall architecture and latent space unchanged, PhysFormer can be extended, in principle, to other high-dimensional systems with unobservable physical quantities. This work thus illustrates a general framework for embedding physical processes directly into generative modeling, providing a practical approach for physically consistent modeling of complex scientific systems.

\bibliographystyle{named}
\bibliography{ijcai26}

\end{document}